\documentclass[conference]{IEEEtran}
\IEEEoverridecommandlockouts
\usepackage{cite}
\usepackage{amsmath,amssymb,amsfonts}
\usepackage{algorithmic}
\usepackage{graphicx}
\usepackage{textcomp}
\usepackage{xcolor, soul}
\usepackage{url}
\usepackage{booktabs}
\usepackage{array}
\newcolumntype{P}[1]{>{\centering\arraybackslash}p{#1}}

\def\BibTeX{{\rm B\kern-.05em{\sc i\kern-.025em b}\kern-.08em
    T\kern-.1667em\lower.7ex\hbox{E}\kern-.125emX}}
\begin{document}

\title{Energy-Rate-Quality Tradeoffs of State-of-the-Art Video Codecs\\
\thanks{This work has been supported by Bristol+Bath Creative Industry Cluster.}
}

\author{\IEEEauthorblockN{Angeliki Katsenou}
\textit{Trinity College Dublin, Dublin, IE}\\
angeliki.katsenou@tcd.ie
\and
\IEEEauthorblockN{Jingwei Mao}
\textit{University of Bristol, Bristol, UK}\\
jingwei.mao.2020@bristol.ac.uk
\and
\IEEEauthorblockN{Ioannis Mavromatis}
\textit{BRIL, Toshiba Europe Ltd., Bristol, UK}\\
ioannis.mavromatis@toshiba-bril.com
}

\maketitle

\begin{abstract}
The adoption of video conferencing and video communication services, accelerated by COVID-19, has driven a rapid increase in video data traffic. The demand for higher resolutions and quality, the need for immersive video formats, and the newest, more complex video codecs increase the energy consumption in data centers and display devices. In this paper, we explore and compare the energy consumption across optimized state-of-the-art video codecs, SVT-AV1, VVenC/VVdeC, VP9, and x.265. Furthermore, we align the energy usage with various objective quality metrics and the compression performance for a set of video sequences across different resolutions. The results indicate that from the tested codecs and configurations, SVT-AV1 provides the best tradeoff between energy consumption and quality. The reported results aim to serve as a guide towards sustainable video streaming while not compromising the quality of experience of the end user.
\end{abstract}


\section{Introduction}
Over the past years, video network traffic is rapidly increasing and currently accounts for the highest Internet-exchanged traffic~\cite{CISCOVNI}. In addition, the recent COVID-19 pandemic contributed to the rapid adoption of digital online services. As a result, live and on-demand video exchange becomes the norm for daily work and leisure activities~\cite{feldmann2021implications}. Popular examples are on-demand streaming platforms (Netflix, Apple TV, HBO, Amazon Prime, etc.) and live video conferencing and collaborative online workspaces (Zoom, Webex, MS Teams, etc.). Furthermore, the accessibility to powerful and affordable devices, and the advances in cloud-computing technologies, enable users to create and share live or on-demand short user-generated content clips over social media/sharing platforms (Instagram, TikTok, YouTube, etc.). 
%Cisco's~\cite{CISCOVNI}, the expectation of video traffic reaching up to 82\% by 2022 is expected to be exceeded. 

Associated with the demands and drivers described above, the content creation and video communications pipelines contribute significantly to global energy consumption. Cloud computing services, data centers, display devices, and video delivery are the main contributors to this increased energy expenditure. While most climate change organizations focus on the transport and energy sectors’ emissions, it is essential to recognize that ICT technologies also generate a considerable carbon emissions footprint~\cite{Greenpeace2015}. Hence, efficiency must improve as technology usage increases if sustainability targets are to be met. According to a recent study by Huawei~\cite{Huawei}, data centers currently consume about 3\% of global electricity. This, however, is expected to rise to over 8\% by 2030, a figure larger than the energy consumption of some nations. While estimates vary, there is a consensus of an impending major global issue. Another critical issue is the energy required from the users to capture, transmit, and display the video data. Recent research has shown that the energy consumed on the user side is much higher than on the provider side~\cite{SchienIEEEComMag2014} given that a single encoding is delivered to thousands of viewers.

While video streaming companies are highly engaged in optimizing their algorithms to offer the highest quality of experience, the energy consumption is not part of this process yet. Each new generation of video codec reduces the amount of data transmitted over the network at the cost of increased computational complexity. A $\sim$50\% efficiency gain of each new codec usually comes with a vast increase in computational complexity~\cite{Ronca, KraenzlerCSVT2021} yielding to significantly increased encoding times. However, decoding has been kept relatively low complying with the requirement for smooth play-outs without rebuffering. With this growth in computational load, video providers, like Netflix, BBC, and others, are working towards assessing the environmental cost and committing towards net zero emissions~\cite{Netflix2021,Schien2021}. 

Various research activities have focused on modeling and predicting the energy consumption at the decoder side~\cite{KaupTCSVT2016, KaupAccess2020}. As a result, tools that analyze the encoding statistics and estimate the decoding energy consumption for H.264/Advanced Video Coding (AVC)~\cite{r:h264}, High Efficiency Video Coding (HEVC)~\cite{r:HEVC}, and Versatile Video Coding (VVC)~\cite{s:VVC1} are currently being developed. Similarly, researchers in~\cite{PakdamanICIP2020, KraenzlerCSVT2021} explore both the encoder and the decoder on the latest VVC standard and compare it against HEVC. 

Challenged by the above, in this work, we investigate the energy, quality, and bitrate tradeoff across different state-of-the-art codecs, particularly, x.265, VVenC/VVdeC, VP9, and Scalable Video Technology AV1 (SVT-AV1). The energy is measured both at the encoder and the decoder side. We selected these production-optimized versions of codecs instead of the reference software implementations as these are usually deployed by the industry. After collecting the quality, rate, and energy statistics, we compare their tradeoffs. Although previous studies have performed codec comparisons in terms of delivered quality and compression effectiveness~\cite{KatsenouICIP2019, KatsenouFrontiers2022}, to the best of our knowledge, this is the first work comparing these codecs with regard to their energy-rate-quality tradeoffs. For the evaluation of the results, a new metric to reflect the energy cost for the required bits is proposed. 
The reported results aim to serve as a guide for the development of sustainable optimization solutions for video compression and streaming algorithms.

The remainder of this paper is organized as follows. Section \ref{sec: Methodology} describes the proposed method and metrics employed. Section \ref{sec: Evaluation} presents the experimental setup, the test configurations, and discussed the results. Finally, conclusions and future work are outlined in Section~\ref{sec: Conclusion}.

\section{Methodology}
\label{sec: Methodology}
In this section, we provide details on the video codecs used, the measurement setup, and the defined metrics.

%?? Consider modelling $E_{enc},E_{dec}$ based on $P_{cpu},fps,NoFrames,Resolution+$ content features? SI? TI? edge entropy?

\subsection{Video Codecs}
\label{ssec: Video Codecs}
 H.264/MPEG-4-AVC \cite{r:h264} was launched in 2004 and remains one of the most widely deployed video coding standards, even though the next generations of standards, H.265/HEVC~\cite{r:HEVC, j:Ohm, b:Wien} and VVC provide enhanced encoding performance~\cite{s:VVC1}. H.265/HEVC was finalized in 2013, and H.266/VVC was released in 2020 with impressive coding gains of over 30\% compared to H.265/HEVC. Besides the activities reported above, there has been increased activity over the past three years in the development of open-source royalty-free video codecs, particularly by the Alliance for Open Media (AOMedia). AOMedia used VP9~\cite{w:VP9}, which was earlier developed by Google, as a basis for AV1~\cite{w:AV1,AV1paper}. AV1 is currently the primary competitor for the current MPEG video coding standards, especially in the context of streaming applications.    
 
  The commercial deployment depends on the hardware and the specifications of the display devices. Moving towards an extended parameter space, namely higher bitdepth (up to 16 bits) and higher spatio-temporal resolutions, the latest standards are expected to roll out to more use-cases and applications. Based on the above standards, optimized implementations of these codecs have been developed and served as part of the FFmpeg software suite~\cite{ffmpeg}. Similarly, SVT-AV1 encoder was developed and optimized for CPU platforms improving the quality-latency tradeoffs for a wide range of video coding applications. It supports multi-dimensional parallelism, multi-pass partitioning decision, multi-stage/multi-class mode decision, and more~\cite{kossentini2020svt}. Shortly after the finalization of VVC, an open-source optimized implementation of the VVC encoder (VVenC) and decoder (VVdeC) for random-access high-resolution video encoding was released~\cite{VVenC,Wieckowski2021}. VVenC was designed to achieve faster runtime than the VVC reference software (VTM). VVenC also supports additional features like multi-threading, rate control and more. 

\begin{table}[ht]
\caption{The video codec software versions and basic configurations.}
\centering 
\begin{tabular}{p{2cm} | p{6cm} }
\toprule
Codec & Command Invocations \\
\midrule
x.265 (ffmpeg4.3) &\texttt{ffmpeg.exe -s \$widthx\$height   -r \$FPS -pix\_fmt \$YUVfmt -i \$input.yuv -c:v libx265 -preset veryfast  -crf \$QP \$output.mp4 } \\
\midrule
VP9 (1.10.0) &\texttt{vpxenc.exe \$input.yuv --width=\$width --height=\$height --verbose --codec=vp9 -o \$output.ivf  --input-bit-depth=\$bd --max-q=\$QP --min-q=\$QP --min-gf-interval=16 --max-gf-interval=16 --kf-min-dist=64 --kf-max-dist=64 --fps=\$FPS/1 --cpu-used=1 --ivf --bit-depth=\$BD}\\

\midrule
SVT-AV1 (0.8.6) &\texttt{SvtAV1EncApp.exe -i \$input.yuv -w \$width -h \$height passes=2 cpu-used=1 
%--enable-stat-report 1 
-b \$output.ivf --fps \$FPS --input-depth \$BD lag-in-frames=\$GoP --keyint \$KI -q \$QP} \\
\midrule
%VVenC/VVdeC (1.0.0)  & \texttt{vvencFFapp.exe -i \$input.yuv --Size \$widthx\$height  --InputBitDepth \$BD -fr \$FPS -b \$output.bin  --OutputBitDepth \$BD -g \$GoP -ip \$KI --QP \$QP --InputChromaFormat 420 --InternalBitDepth \$BD} \\

VVenC/VVdeC (1.0.0)  & \texttt{vvencapp.exe -s \$widthx\$height  -r \$FPS -c yuv420\_10 -i \$input.yuv  --preset medium  -q \$QP -o \$output.bin}\\
\bottomrule
\end{tabular} 
\label{tab:codecs}
\end{table}

\subsection{Measurements}
\label{ssec: Measurements}
We propose to measure the energy consumption on both the encoder and the decoder ends. The encoding side is a good representation of the energy consumption at the video provider's side (e.g. in data centers), while the decoding end reflects directly to the decoding at the end-user devices (typically mobile devices, laptops, or TVs). We formulate and perform two basic measurements. The first power measurement is performed during encoding, $P_{enc}$, and the second during decoding, $P_{dec}$. A third power measurement $P_{idle}$ quantifies the idle mode of our system. The energy consumption during the encoding and decoding is derived from the measured power minus the idle time measurements. Thus, over an observed time interval, the encoding and decoding energy is obtained by:
\begin{equation}
    E_{enc}=\int^{T_{enc}}_{t=0} P_{enc}(t) dt-\int^{T_{enc}}_{t=0} P_{idle}(t) dt
\end{equation}
\begin{equation}
    E_{dec}=\int^{T_{dec}}_{t=0} P_{dec}(t) dt-\int^{T_{dec}}_{t=0} P_{idle}(t) dt
\end{equation}
where $T_{enc}$ and $T_{dec}$ are the encoding and decoding times.

Our performance investigation is based on the integrated power meter in Intel CPUs, the Running Average Power Limit (RAPL)~\cite{RAPL2018}. This tool is used in other similar research activities (e.g.,~\cite{KaupTCSVT2016,KaupVCIP2020}) and accurately measures the power demand of the CPU, the DRAM, and the whole integrated circuit at 100ms intervals. Background processes of the operating system can skew our measurements. Therefore, we started our experimentation with a small-scale study assessing the energy consumption at an idle state, $E_{idle}$.  
Later, by profiling the power consumption during the encoding and decoding process, we can assess the energy requirements for both processes. We repeat until the confidence intervals of the distribution of the measurements are tight validating their precision. For the encoding process, a smaller number of encoding iterations was required to converge, while for the decoding a greater number of decoding loops was necessary. This is attributed to the significant difference in the time duration of the encoding and decoding processes.

\subsection{Metrics}
\label{ssec: Metrics}
\subsubsection{Quality performance}
To assess the quality of the encoded test sequences, we selected full reference metrics typically used over the last years in the video technology research community: the Peak Signal to Noise Ratio (PSNR) averaged over all color components (YUV) and Video Multi-Method Assessment Fusion (VMAF)~\cite{VMAFblog}. The latter exhibits a higher correlation with perceptual quality.

\subsubsection{Compression performance}
The performance of video coding algorithms is usually assessed by comparing their rate-distortion (or rate-quality) performance on various test sequences. Objective quality metrics or subjective opinion measurements are normally employed to assess compressed video quality, and the overall Rate-Quality (RQ) performance difference between codecs can be then calculated using Bj{\o}ntegaard Delta (BD) measurements \cite{r:Bjontegaard} on objective metrics. We computed the BD metrics for the PSNR-Rate and VMAF-Rate curves. 

\subsubsection{Energy performance}
% To compare the energy performance across different codecs, we adjusted the BD metric on the Rate-Energy (RE) curves. Particularly, the average energy difference between two RE curves is approximated by the difference between the integrals of the fitted RE curves divided by the integration interval. : %We are also using the average energy consumption for a subset range of QPs that commonly results in bitrates within the streaming  range.
% \begin{equation}
%     \Delta E = \dfrac{1}{R_h - R_l} \int^{R_h}_{R_l} (\hat{E}_{t}(r) - \hat{E}_{ref})(r) dr \;
% \end{equation}
% where $\hat{E}$ denotes the estimated energy at the rate point $r$ when using the piece-wise cubic hermitian interpolating polynomial. 

To compare the energy performance across different codecs, we need to express the required energy as a function of the bitrate. By observation, the Rate-Energy (RE) curves both for encoding and decoding are very close to linear (see Fig.~\ref{fig: averageREs}). Therefore, based on the slope of the RE line, we define another metric, the Energy-to-Bitrate Ratio (EBR). EBR expresses the required energy expenditure for different compression levels. The lower the slope value (close to zero), the lower the EBR between the tested compression levels and the more energy-efficient the compression technology. In order to compute the slope of the RE curves, we first fitted the RE points into a linear model, namely
\begin{equation}
    \Tilde{E} = \alpha R + \beta \; ,
\end{equation}
where $\alpha, \beta \in \mathbb{R}^+$. $\Tilde{E}$ is the estimated energy either for encoding or decoding. It is well known that in first-order polynomials $\alpha$ expresses the slope. Thus, EBR values are equal to $\alpha$. The close-to-linearity behavior of the RE curves is confirmed through testing on the dataset described below. All R-squared values are higher than 0.92 indicating a very good fit.

\section{Evaluation}
\label{sec: Evaluation}
In this section, we describe the test sequences employed for the evaluation, the basic codec configurations, and report on the findings from our experiments.

\subsection{Test Data}
\label{ssec: TestData}
\begin{figure}[htbp]
\centerline{\includegraphics[width=0.85\linewidth]{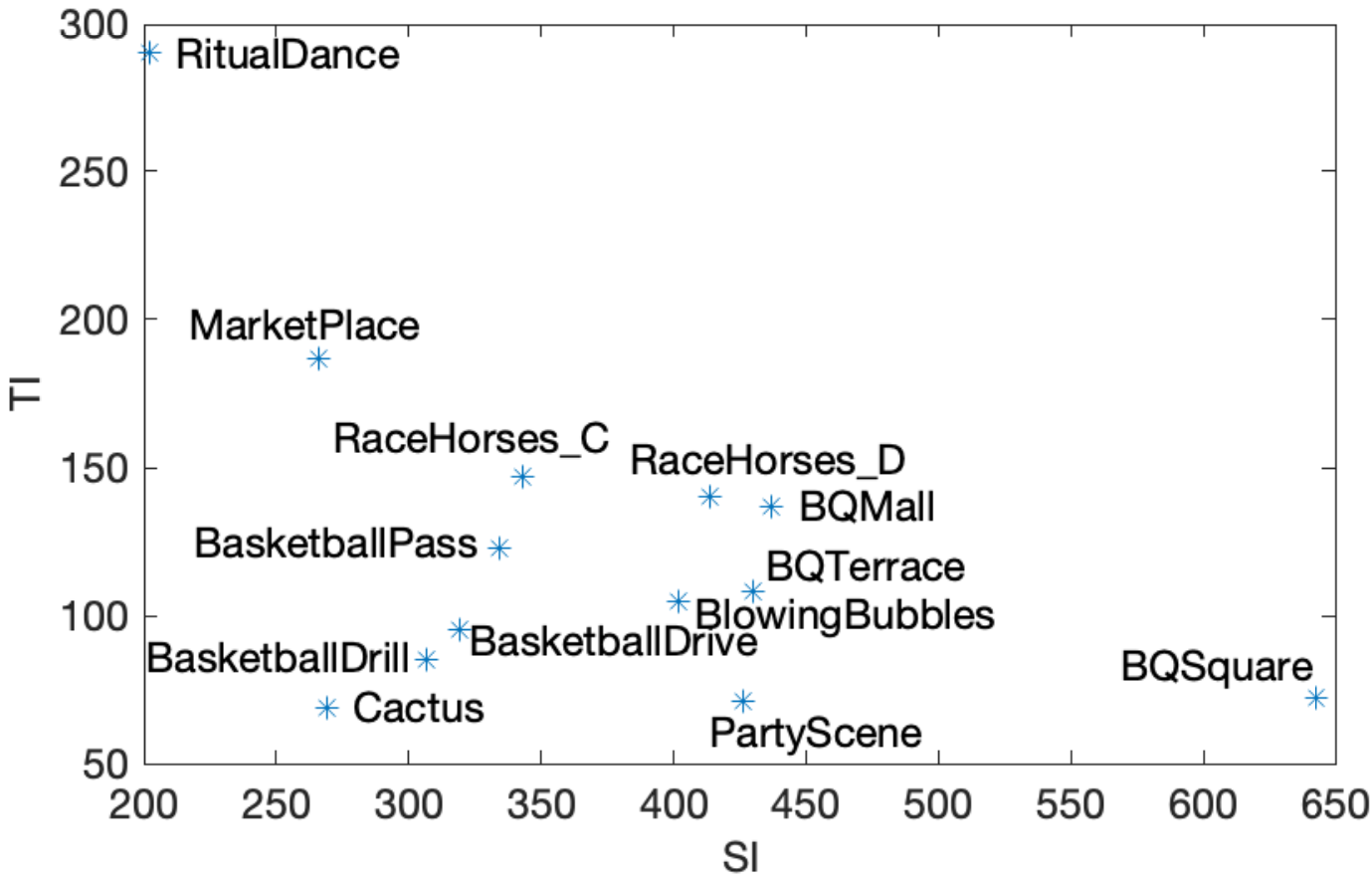}}
\vspace{-1em}
\caption{SI against TI for the test sequences.}
\vspace{-1em}
\label{fig: SIvsTI}
\end{figure}

The selection of test content is important as compression is content-dependent and  should provide a diverse and representative coverage of the video parameter space. For our experiments, we selected the SDR CTC test sequences~\cite{CTC2019} reported in Table~\ref{tab:TestSeqs}. These sequences have been used for many video codec evaluations, as they cover a typically used range of spatial resolutions \{Class D: 416$\times$240, Class C: 832$\times$480, Class B: 1920$\times$1080, Class A: 3840$\times$2160\}, frame rates -from 30 to 60-, and bit depths -from 8 to 10 bits. Besides this, the content also covers a representative range of spatial and temporal characteristics, as indicated by the spatial and temporal information~\cite{Barman2019,itup910} scattered in Fig.~\ref{fig: SIvsTI}. In the presented results, we have considered the sequences with resolutions up to 1920x1080 (class B). These results are adequate to convey the energy consumption trends.

\begin{table}[ht]
\caption{Test Sequences and basic characteristics.}
\centering 
\begin{tabular}{c | l | c | P{1.5cm} | P{1.5cm}}
\toprule
Class	& Name & No Frames & Frame Rate (\$FPS) & Bitdepth (\$BD)\\
\midrule
B& MarketPlace& 600& 60& 10\\
B& RitualDance& 600& 60& 10\\
B& Cactus& 500& 50& 8\\
B& BasketballDrive& 500& 50& 8\\
B& BQTerrace& 600& 60& 8\\
\midrule
C& RaceHorces& 600& 30& 8\\
C& BasketballDrill& 500& 50& 8\\
C& PartyScene& 500& 50& 8\\
C& BQMall& 300& 60& 8\\
\midrule
D& RaceHorces& 600& 30& 8\\
D& BasketballPass& 500& 50& 8\\
D& BlowingBubbles& 500& 50& 8\\
D& BQSquare& 300& 60& 8\\
\bottomrule
\end{tabular} 
\label{tab:TestSeqs}
\vspace{-1em}
\end{table} 

\subsection{Settings}
Table~\ref{tab:codecs} summarises the codec and the software versions deployed for our study. 
The selected codecs have a different range of quantization parameters: $[0-51]$ for HEVC/VVenC and $[0-63]$ for VP9/SVT-AV1. For the encoding points, we selected for HEVC/VVenC the \texttt{\$QP} values from the JVET recommended range~\cite{CTC2019}, $\{22,27,32,37\}$. For VP9/SVT-AV1 codecs, we performed a linear mapping on the range and rounded to the nearest integer. This resulted in $\{27,33,40,46\}$. For x.265, we selected the fastest preset in order to create an anchor that would represent an efficient codec that has low energy consumption. Note that the different codec configurations are switching on and off coding tools with a direct impact on the computational complexity and, thus, on the energy consumption. Here, we examine only a subset of the available configurations.
All experiments were executed on the same workstation with a Hexacore Intel Comet Lake-S CPU at 3300MHz and 64GB RAM. More technical details can be found in the project page~\footnote{\url{https://angkats.github.io/Sustainability-VideoCodecs/}}.

\subsection{Results}
\label{ssec: Results}
\begin{figure}[htbp]
\begin{minipage}[b]{.49\linewidth}
    \centerline{\includegraphics[width=\linewidth]{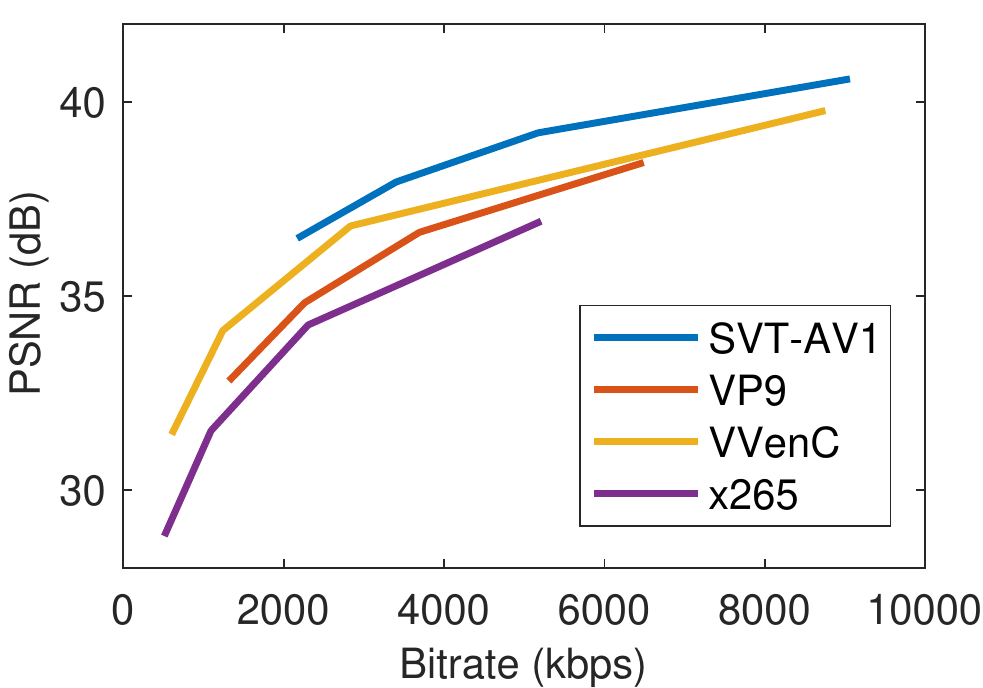}}

    \centerline{\scriptsize{(a) PSNR (dB) vs Bitrate (kbps).}}
\end{minipage}
\hfill
\begin{minipage}[b]{.49\linewidth}
    \centerline{\includegraphics[width=\linewidth]{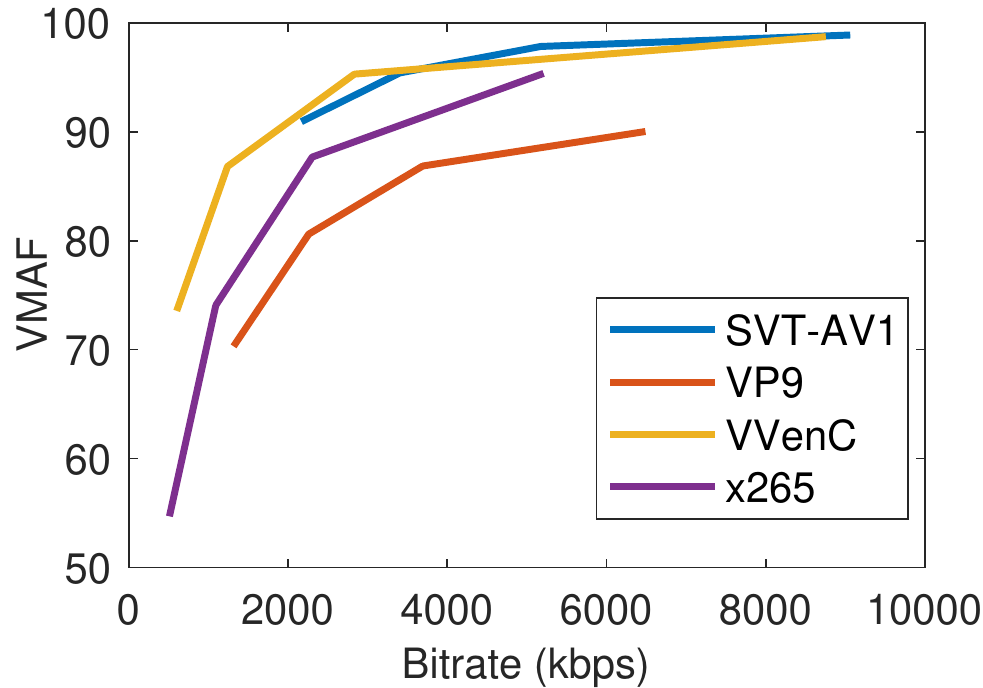}}
    
    \centerline{\scriptsize{(b) VMAF vs Bitrate (kbps).}}
\end{minipage}
\caption{Average Rate-Quality Curves.}
\vspace{-1em}
\label{fig: averageRQs}
\end{figure}

The computed performance metrics, BD and EBR, for the tested codecs are reported in Table~\ref{tab:Metrics}. For the BD metrics computation, the x.265 was considered the anchor codec. The BD metrics were computed for both types of RQ curves, Rate-PSNR and Rate-VMAF. Only the BD-PSNR (in dB) and BD-VMAF are reported in pairs. We did not include the BD-Rate results in this table, because differences in quality
scales in most cases are so big that the extrapolation across the bitrate axis to compute the integral differences would not be accurate. It is clear from the BD-PSNR values, that SVT-AV1 offers more gains over x.265 compared to the other codecs. On the other hand, according to BD-VMAF VVenC/VVdeC offers on average the highest perceptual gains. Overall, VVenC/VVdeC seems to be offering a good tradeoff for the achieved quality, especially at low bitrate ranges. These results are confirmed by the plots of the average RQ curves in Fig.~\ref{fig: averageRQs}.

\begin{figure}[htp]
\begin{minipage}[b]{.49\linewidth}
    \centerline{\includegraphics[width=\linewidth]{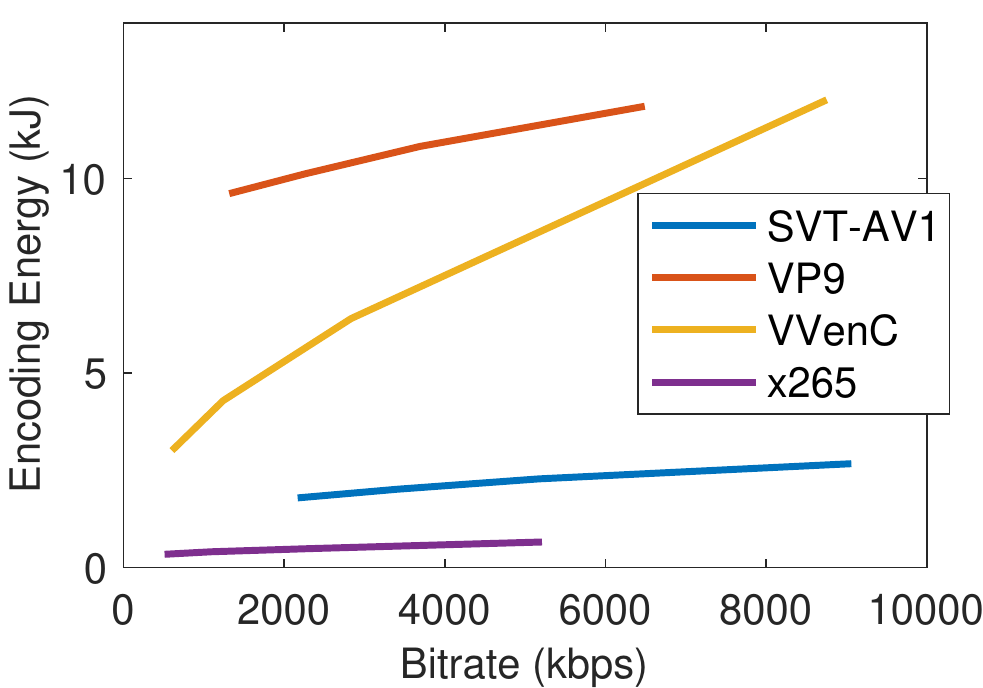}}
    
     \centerline{\scriptsize{(a) Encoding Energy vs Bitrate.}}
\end{minipage}
\hfill
\begin{minipage}[b]{.49\linewidth}
    \centerline{\includegraphics[width=\linewidth]{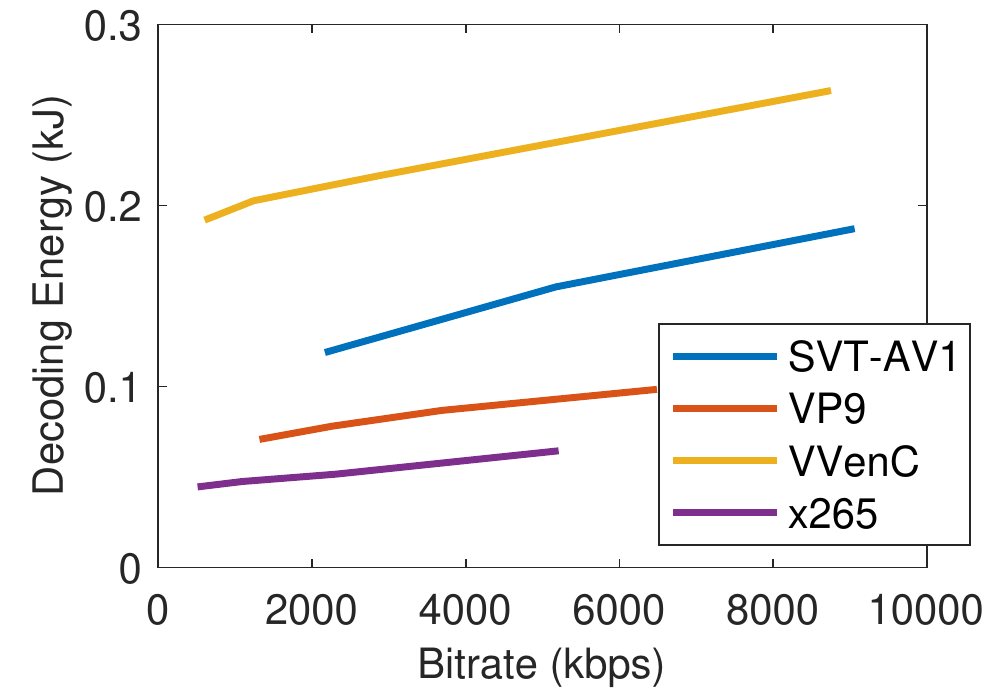}}
    
    \centerline{\scriptsize{(b) Decoding Energy vs Bitrate.}}
\end{minipage}
\caption{Average Rate-Energy Curves.}
\vspace{-1.5em}
\label{fig: averageREs}
\end{figure}

The observed improved compression performance of SVT-AV1 and VVenC/VVdeC in terms of quality comes at the cost of higher complexity and, thus, energy consumption. This is confirmed by the RE curves in Fig.~\ref{fig: averageREs} for both encoding and decoding. It is also noticeable from these figures and the EBR values in Table~\ref{tab:Metrics} that the two latest codecs, SVT-AV1 and VVenC/VVdeC, have an almost equivalent slope in decoding, although VVdeC requires more energy for decoding. Regarding encoding, SVT-AV1 is performing significantly better than AV1 with EBR\_enc values comparable to those of x.265. It is also worth mentioning that although VP9 demonstrates the second best EBR\_dec and average decoding energy consumption, its encoding energy consumption is the highest on average compared to the other codecs. 

\begin{figure}[htp]
\begin{minipage}[b]{.49\linewidth}
    \centerline{\includegraphics[width=\linewidth]{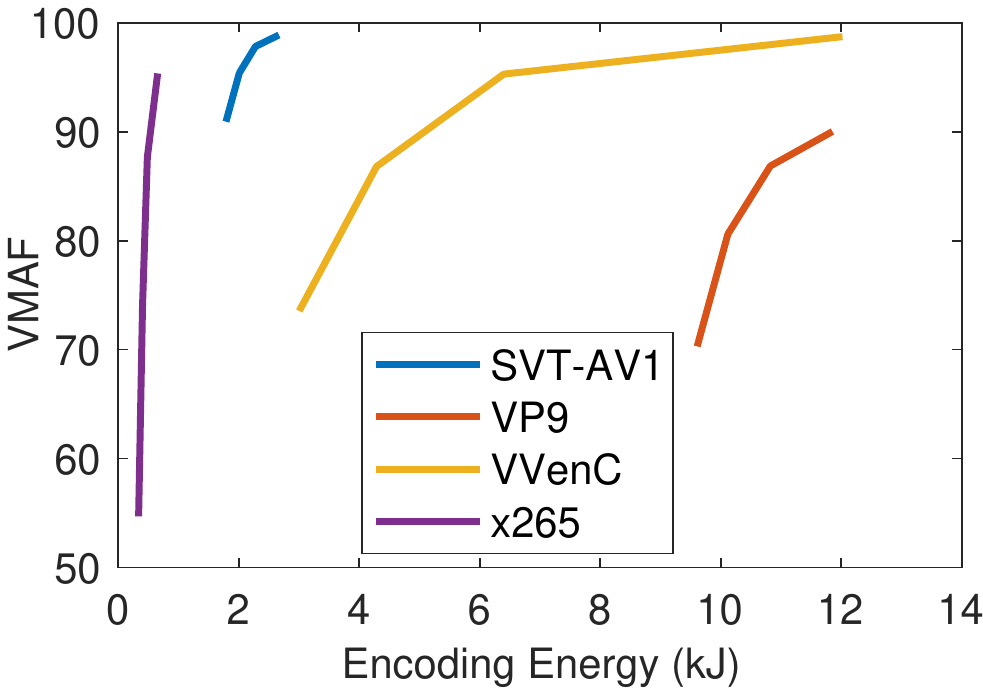}}
    
    \centerline{\scriptsize{(a) VMAF vs Encoding Energy.}}
\end{minipage}
\hfill
\begin{minipage}[b]{.49\linewidth}
    \centerline{\includegraphics[width=\linewidth]{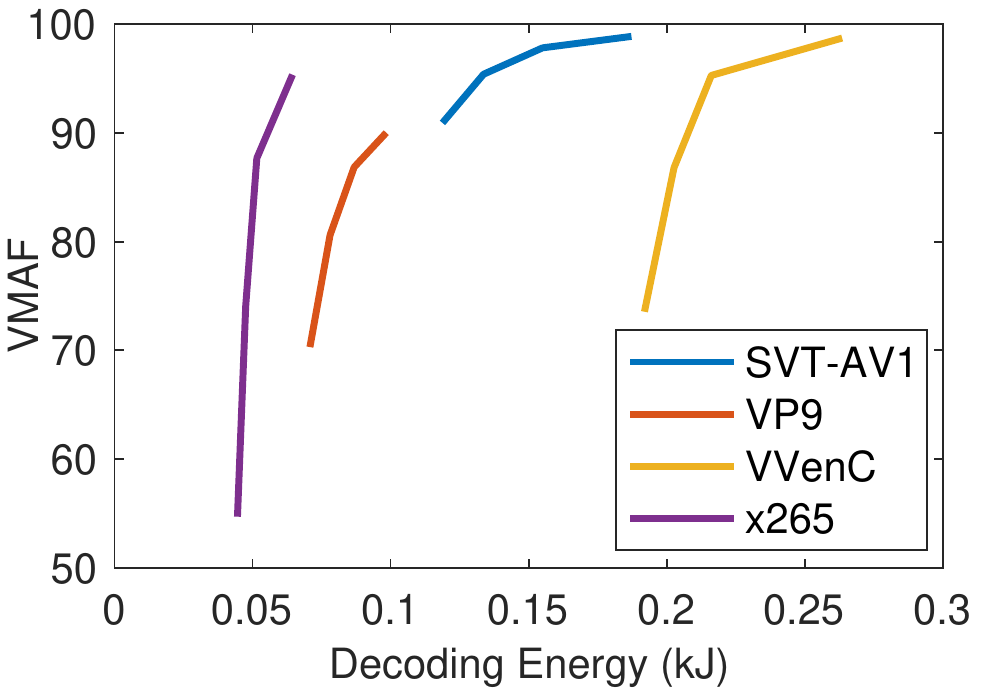}}
    
     \centerline{\scriptsize{(b) VMAF vs Decoding Energy.}}
\end{minipage}

\begin{minipage}[b]{.49\linewidth}
    \centerline{\includegraphics[width=\linewidth]{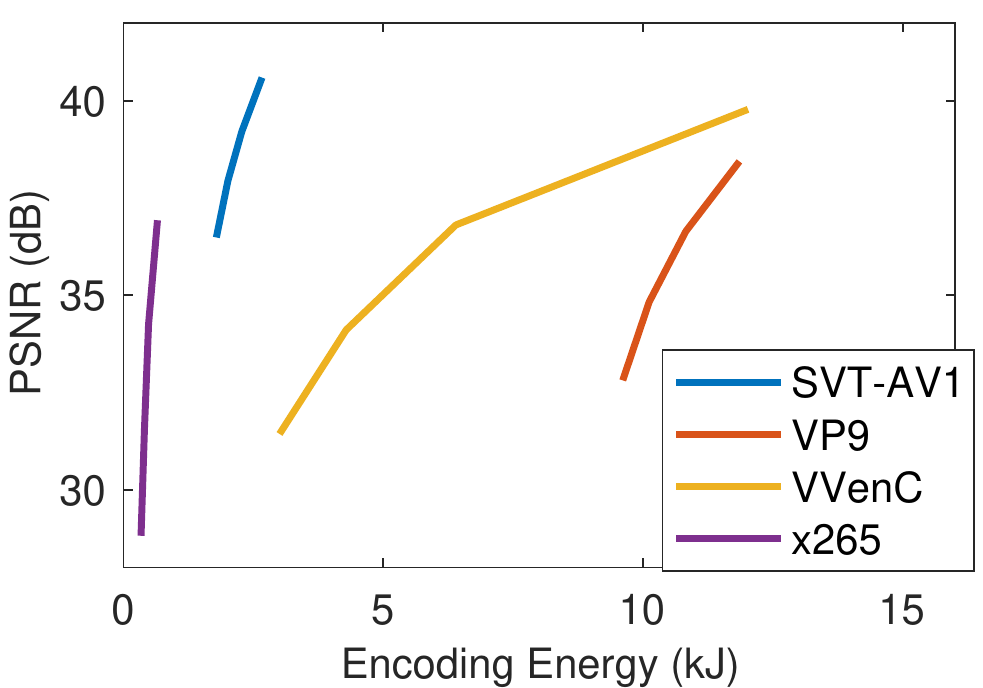}}
    
    \centerline{\scriptsize{(c) PSNR vs Encoding Energy.}}
\end{minipage}
\hfill
\begin{minipage}[b]{.49\linewidth}
    \centerline{\includegraphics[width=\linewidth]{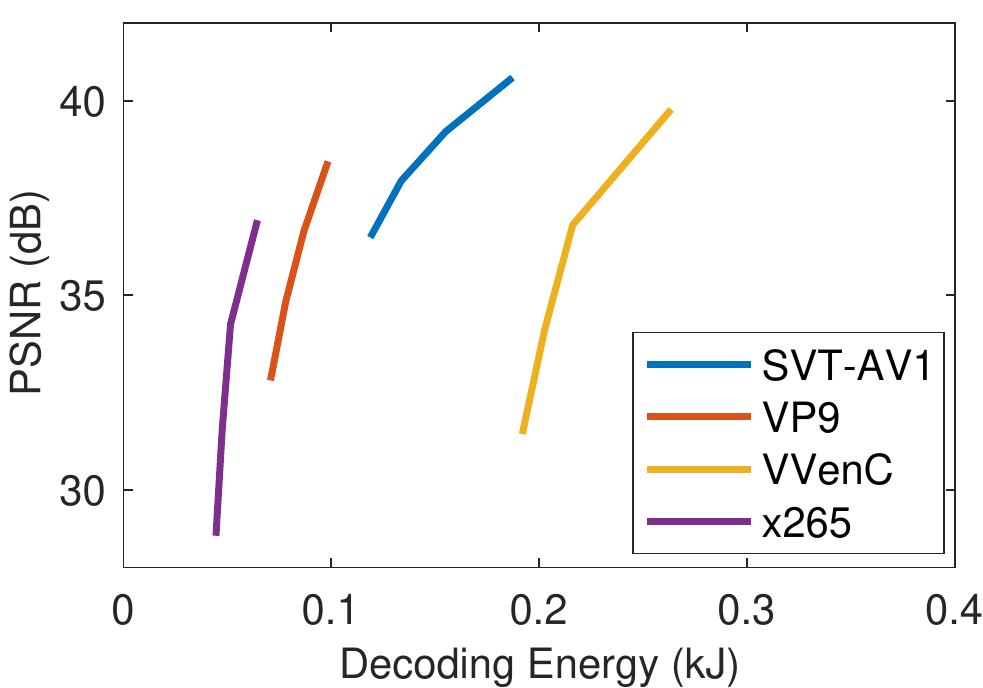}}
    
     \centerline{\scriptsize{(d) PSNR vs Decoding Energy.}}
\end{minipage}

\caption{Average Quality-Energy Curves.}
\vspace{-1.5em}
\label{fig: averageQEs}
\end{figure}

Another interesting view of the tradeoffs between quality and encoding/decoding energy can be derived from Fig.~\ref{fig: averageQEs}, where the average PSNR and VMAF are plotted against the encoding and decoding energy required. It is observed from these plots that SVT-AV1 appears to offer the best tradeoff on the encoding side in terms of quality and required energy, as it achieves on average a very high quality (over 90 in terms of VMAF). 

All these results reflect the expected energy consumption of these four codecs in a peer-to-peer scenario and for the specific codec configurations. It is expected that the codecs could behave differently under different settings.% In order to make this study complete, we should extend the set of configurations. %Besides this, in order to complete the 

\begin{table*}[ht]
\caption{Performance Metrics of RQ and RE curves. The BD metrics have been computed using x.265 as the anchor. The EBR metric is computed both at the encoder and the decoder (EBR\_enc, EBR\_dec).}
\centering 
\begin{tabular}{l| l| c c c | c c c c}
\toprule
	& &\multicolumn{3}{|c|}{[BD-PSNR (dB), BD-VMAF]} & \multicolumn{4}{|c}{[EBR\_enc, EBR\_dec]}\\
Class	& Sequence &SVT-AV1 & VP9 & VVenC/VVdeC & SVT-AV1 & VP9 & VVenC/VVdeC & x.265 \\
\midrule
B& MarketPlace&-2.11, -3.03 & -1.19, -1.73 & -1.35, -8.08 &0.145, 0.011 & 0.491, 0.006 & 1.325, 0.010 & 0.067, 0.004\\
B& RitualDance&-2.51, -2.40 & -0.91, 0.44 & -0.80, -4.08 &0.121, 0.011 & 0.614, 0.005 & 0.900, 0.008 & 0.073, 0.006\\
B& Cactus&-1.68, -2.59 & -0.40, 1.95 & -1.27, -7.52 &0.131, 0.010 & 0.602, 0.005 & 1.150, 0.009 & 0.059, 0.004\\
B& BasketballDrive&-2.09, -3.13 & -0.46, 1.05 & -1.25, -7.43 &0.157, 0.009 & 0.351, 0.006 & 1.161, 0.008 & 0.083, 0.003\\
B& BQTerrace&-2.50, -3.84 & -1.02, -1.75 & -1.90, -8.17 &0.127, 0.009 & 0.072, 0.006 & 1.253, 0.011 & 0.078, 0.005\\
\midrule
C& RaceHorces&-2.07, -3.09 & 0.02, 0.16 & -2.15, -7.97 &0.106, 0.007 & 0.461, 0.004 & 0.941, 0.009 & 0.052, 0.003\\
C& BasketballDrill&-2.56, -2.18 & -0.49, 92.53 & -1.94, -6.76 &0.137, 0.009 & 0.313, 0.004 & 1.188, 0.008 & 0.080, 0.005\\
C& PartyScene&-2.44, -3.78 & -0.16, 1.13 & -2.39, -9.94 &0.080, 0.007 & 0.243, 0.004 & 0.766, 0.008 & 0.036, 0.004\\
C& BQMall&-1.92, -2.75 & -1.38, -3.37 & -1.16, -5.94 &0.072, 0.006 & 0.226, 0.003 & 0.682, 0.007 & 0.036, 0.003\\
\midrule
D& RaceHorces&-2.21, -4.17 & -0.86, -2.5 & -1.87, -8.23 &0.072, 0.006 & 0.206, 0.003 & 0.577, 0.005 & 0.046, 0.004\\
D& BasketballPass&-2.49, -3.78 & -0.23, 0.94 & -2.04, -7.9 &0.093, 0.009 & 0.402, 0.004 & 0.604, 0.005 & 0.036, 0.004\\
D& BlowingBubbles&-3.55, -3.81 & -0.76, -0.08 & -2.35, -6.02 &0.103, 0.010 & 0.392, 0.004 & 0.691, 0.008 & 0.042, 0.007\\
D& BQSquare&-2.04, -2.97 & -1.22, -3.52 & -2.05, -5.84 &0.077, 0.008 & 0.224, 0.003 & 0.613, 0.005 & 0.040, 0.003\\
\midrule
\multicolumn{2}{c|}{\textbf{Average}} &  \textbf{-2.32}, -3.19&   -0.69,  6.55   &1.73, \textbf{-7.22} & 0.109, 0.009&    0.354, 0.005&    0.912, 0.008&    \textbf{0.056, 0.004}\\ 
\bottomrule
\end{tabular} 
\label{tab:Metrics}
\vspace{-1em}
\end{table*}

\section{Conclusion}
\label{sec: Conclusion}
In this paper, we presented a study on the energy consumption of four state-of-the-art codecs, SVT-AV1, VP9, VVenC, and x.265, that are optimized to be used in production. The experimental setup was built based on the codecs CTCs and using a video dataset with sequences with spatial resolution from 480p to 1080p. For the evaluation of the results,
a new metric to reflect the energy cost for the required bits was proposed. From the results acquired with the specific set of coding configurations explored, SVT-AV1 offers the best quality-bitrate-energy tradeoff compared to the other codecs. On the other hand, for low-energy solutions, x.265 seems to be the best choice at the cost of lower video quality on average.

Future work will include further experimentation with the different codec configurations. Moreover, we plan to extend this use case to include an estimation of the networking energy consumption to study the impact of the audience size on the energy consumed by the end user (decoding energy). Furthermore, the concept of energy-driven codec selection associated with the content and audience size under the constraint of maintaining a high user experience will also be explored.

\bibliographystyle{IEEEbib}
\bibliography{refs}

\end{document}